\documentclass[10pt]{article}
\usepackage[utf8]{inputenc}
\usepackage{amsmath,amssymb,graphicx}
\usepackage{hyperref}
\usepackage[T1]{fontenc}
\usepackage{mathpazo}
\usepackage[font=small,labelfont=bf]{caption}
\usepackage{longtable}
\usepackage{verbatim}
\usepackage{amsfonts}
\usepackage{cite}
\usepackage{float}
\usepackage[utf8]{inputenc}

\usepackage[table,usenames,dvipsnames]{xcolor}
\definecolor{darkerblue}{rgb}{0.2,0.2,0.5}

\usepackage{afterpage}
\usepackage[utf8]{inputenc}

\usepackage{enumerate}

\newcommand{\bear}{\begin{array}}
\newcommand{\ear}{\end{array}}

\newcommand{\beq}{\begin{eqnarray}}
\newcommand{\eeq}{\end{eqnarray}}
\newcommand{\beqa}{\begin{eqnarray}}
\newcommand{\eeqa}{\end{eqnarray}}

\def\OMIT#1{{}}
\newcommand{\lsim}{\mathrel{\rlap{\lower4pt\hbox{\hskip1pt$\sim$}}
    \raise1pt\hbox{$<$}}}         
\newcommand{\gsim}{\mathrel{\rlap{\lower4pt\hbox{\hskip1pt$\sim$}}
    \raise1pt\hbox{$>$}}}         

\usepackage{tikz}
\usetikzlibrary{trees}
\usetikzlibrary{decorations.pathmorphing}
\usetikzlibrary{decorations.markings}
\tikzset{
    photon/.style={decorate, decoration={snake}, draw=black},
    wino/.style={draw=redwine},    
    electron/.style={draw=black, postaction={decorate},
        decoration={markings,mark=at position .55 with {\arrow[draw=black]{>}}}},
    scalar/.style={draw=black, dashed,postaction={decorate},
        decoration={markings,mark=at position .55 with {\arrow[draw=black]{>}}}},
    gluon/.style={decorate, draw=black,
        decoration={coil,amplitude=4pt, segment length=5pt}}
}

\usepackage{titlesec}

\title{\bf \color{ForestGreen} Ultralight Repulsive Dark Matter and BEC}

\author{JiJi Fan \\
{\em Department of Physics, Brown University, Providence, RI 02912}}

\begin{document}
\maketitle

\begin{abstract}
Ultralight scalar dark matter with mass at or below the eV scale and pressure from repulsive self-interaction could form a Bose-Einstein condensate in the early Universe and maybe in galaxies as well.  It has been suggested to be a possible solution to the cusp/core problem or even to explain MOND phenomenology.
In this paper, I initiate a study of possible self-interactions of ultralight scalar dark matter from the particle physics point of view. To protect its mass, the scalar dark matter is identified as a pseudo Nambu-Goldstone boson (pNGB). Quite a few pNGB models with different potentials such as the QCD axion and the dilaton lead to attractive self-interactions. Yet if an axion is a remnant of a 5D gauged $U(1)$ symmetry, its self-interactions could be repulsive provided the masses and charges of the 5D matter contributing to its potential satisfy certain constraints. Collective symmetry breaking could also lead to a repulsive self-interaction yet with too large a strength that is ruled out by Bullet Cluster constraints. I also discuss cosmological and astrophysical constraints on ultralight repulsive dark matter in terms of a parametrization motivated by particle physics considerations. 
\end{abstract}

\tableofcontents

\section{Introduction}
\label{sec:intro}

The existence of dark matter is well established through cosmological and astrophysical observations. Yet little is known about its nature and interactions beyond gravity. This lack of knowledge leads to a wide range of theoretical speculations with dramatically different experimental consequences. 
In the zoo of dark matter models, one distinctive interesting direction is that (cold) bosonic dark matter is ultralight with mass around or below the eV scale and behaves as a single coherent wave instead of individual particles in galaxies. The dark matter may form a Bose-Einstein condensate (BEC) with a long range correlation. 
 This scenario has been suggested as one possible solution to the cusp/core problem (e.g.~\cite{Flores:1994gz, Moore:1994yx, Burkert:1995yz, Moore:1999gc, Salucci:2000ps, Strigari:2008ib, Donato:2009ab, deBlok:2010}). There are two classes of proposals: fuzzy dark matter with mass $\sim 10^{-22}$ eV and negligible self-interactions~\cite{Hu:2000ke} and repulsive dark matter with mass between eV and $10^{-22}$ eV and a (small) repulsive quartic self-interaction~\cite{Tkachev:1986tr, Goodman:2000tg, Peebles:2000yy}. In the first case, the dark matter mass is fixed and the corresponding de Broglie wavelength is of order kpc. The quantum pressure opposed to localizing particles could stabilize against gravitational collapse and support a halo core. In the second case, the dark matter mass could be in a wide range. The quantum pressure is not sufficient and additional pressure from a repulsive self-interaction is needed to balance against gravity and support a kpc sized core. More recently, a new class of ultralight repulsive dark matter has been proposed to modify the dark matter properties at the galactic scale and explain MOND phenomenology, e.g., the baryonic Tully-Fischer relation~\cite{Berezhiani:2015pia, Berezhiani:2015bqa}. In this proposal, the ultralight bosonic dark matter has a repulsive sextic self-interaction, leading to a different equation of state compared to the one in the quartic interaction case. 

There have already been quite a few studies on the cosmological and astrophysical phenomenology of ultralight scalar dark matter. On the particle/string theory side, it is known that ultralight axions with a wide range of masses could arise and serve as dark matter candidates~\cite{Svrcek:2006yi, Arvanitaki:2009fg, Kim:2015yna}. Yet there are very few studies on the possible self-interactions of ultralight scalar dark matter. As mentioned in the brief review above, self-interactions of ultralight dark matter, in particular, the repulsive ones, may lead to highly nontrivial galactic dynamics. It is thus the purpose of this paper to initiate a study from the particle physics point of view possible (repulsive) self-interactions of ultralight scalar dark matter. I will discuss the difficulties to construct a reasonable particle physics model, present a working example and show that particle physics considerations restrict viable parameter space one should study for cosmological and astrophysical consequences.

Dark BEC has also been studied in the context of self-interacting dark matter scenario, which is another proposal to tackle the cusp/core problem (beginning with~\cite{Spergel:1999mh, Dave:2000ar}, for more recent progress, see~\cite{Rocha:2012jg, Peter:2012jh, Zavala:2012us}). In that case, any cuspy feature at the galactic center is smoothed out by dark matter collisions. 
The bosonic dark matter particles are much heavier than the eV scale~\cite{Tulin:2013teo} and the BEC they could form leads to more compact objects such as a dark star~\cite{Bramante:2013hn, Eby:2015hsq, Soni:2016gzf}. 

The paper is organized as follows: in Sec.~\ref{sec:conditions}, I will present a general parametrization of the potential of ultralight scalar dark matter motivated by particle physics considerations and discuss some minimal necessary conditions for the formation of dark BEC. In Sec.~\ref{sec:model}, I discuss various difficulties of constructing an ultralight dark matter model with repulsive self-interactions. I'll present one working example and several instructive counter-examples. In Sec.~\ref{sec:cos}, I discuss some cosmological and astrophysical implications. I'll conclude and discuss future directions in Sec.~\ref{sec:con}.

\section{Parametrization and Necessary Conditions for Dark Matter BEC}
\label{sec:conditions}

\subsection{Parametrization of the Scalar potential}
\label{sec:par}

In general, the perturbative relativistic Lagrangian of an ultralight real scalar $\phi$ could be parametrized as
\beq
{\cal L}_\phi= \frac{1}{2} (\partial \phi)^2 - \left( \frac{1}{2} m^2 \phi^2 + \frac{\lambda_3}{3!} \frac{m^2}{f} \phi^3 + \frac{\lambda_4}{4!} \frac{m^2}{f^2} \phi^4 \right)+\cdots,
\label{eq:scalar}
\eeq 
which should be taken as an effective field theory description valid below the energy scale $4\pi f$ with $f \gg m$. $\lambda_3$ and $\lambda_4$ are dimensionless coefficients, whose natural values are of order one. For the axion case (or a Goldstone boson associated with the breaking of a compact group), $\lambda_3$ is zero. Yet for the other models, e.g., scalar dark matter associated with a non-compact flat direction in the field space, $\lambda_3$ is generically non-zero. All the models in Sec.~\ref{sec:model} lead to Eq.~\ref{eq:scalar}. Fuzzy dark matter with mass $\sim 10^{-22}$ eV and negligible self-interactions should be taken as the limit of a large $f$, which I will quantify in Sec. 4, instead of from a free particle model. 

I will only keep renormalizable interactions. Higher dimensional operators including higher-derivative interactions will be suppressed by more powers of $m/f$ and are usually subdominant. Even in special cases that higher dimensional operators are dominant at tree level, one should worry about renormalizable operators induced through loops with insertions of higher-dimensional operators. It has been suggested in Refs \cite{Berezhiani:2015pia, Berezhiani:2015bqa} that if the leading interactions are sextic terms, the model may lead to MOND dynamics. This is a very intriguing proposal yet more work is needed to check whether there is a reasonable particle physics model with naturally suppressed lower dimensional operators and leading interactions from high-dimensional operators. In this work, I will focus on the potential in Eq. \ref{eq:scalar}.

In the non-relativistic (NR) limit, one could write 
\beq
\phi = \frac{1}{\sqrt{2m}} \left(e^{-imt} \psi(t, \vec{x}) + e^{imt} \psi^*(t, \vec{x}) \right). 
\label{eq:nrpr}
\eeq
Notice that in this parametrization, $\psi$ has mass dimension $3/2$. 
Plugging this back to the kinetic and mass terms of the relativistic Lagragian, one could obtain the NR kinetic term. Matching the three and four point functions in the relativistic and NR theories, one could obtain the Wilson coefficients of the interaction terms in the NR Lagrangian. The final result of the leading-order NR Lagrangian is 
\beq
{\cal L}_{\rm NR} &=& \frac{i}{2} \left(\dot{\psi} \psi^* - \psi \dot{\psi}^* \right) - \frac{\left| \nabla \psi \right|^2}{2m}  - \frac{\lambda_4^{\rm eff}}{16 f^2} \left|\psi\right|^4, \nonumber \\
\lambda_4^{\rm eff}&=&\lambda_4 - \frac{5}{3}\lambda_3^2.
\label{eq:effectivecoupling}
\eeq
Notice that there is no cubic interaction in the NR limit. This could be understood easily as in the NR limit, a single $\phi$ (almost at rest) could not create two $\phi$'s (almost at rest). Also notice that the quartic coupling in the NR limit is shifted from its relativistic value when one integrates out heavy modes with energy $\approx m$. A cubic interaction in the relativistic Lagragian always yields a {\it negative} contribution to the quartic interaction in the NR limit. Whether the particles have repulsive or attractive self-interactions depends on the sign of $\lambda_4^{\rm eff}$: when $\lambda_4 > \frac{5}{3} \lambda_3^2$, the self-interaction is repulsive; otherwise, it is attractive.

\subsection{Necessary Conditions for Dark Matter BEC}
\label{sec:nconditions}

If the dark matter particles could condense, the particle's de Broglie wavelength has to be larger than the inter-particle spacing so that the wave functions of individual particles overlap with each other:
\beq
\lambda_{dB} = \frac{2 \pi}{m v} > \sqrt[3]{\frac{m}{\rho}} \Rightarrow m < 1\,{\rm eV} \left(\frac{10^{-3}}{v}\right)^{3/4}\left(\frac{\rho}{1.3 \times 10^{-6}\, {\rm GeV}/{\rm cm}^3}\right)^{1/4},
\eeq
where $1.3 \times 10^{-6}$ GeV/cm$^3 \approx 2.8 \times 10^{11} \Omega_{\rm DM} h^2 M_\odot/{\rm Mpc}^3$ is the average density of dark matter in the Universe. In the early Universe, the density will be higher and this condition could be satisfied more easily. In galaxies, dark matter density will also be higher than the average and has a nonuniform distribution. Thus the inequality above should just be taken as an order of magnitude estimate. 
On the other hand, if the de Broglie wavelength is above the galactic scale, the large scale structure could be modified. To avoid that, I take the de Broglie wavelength to be around or below kpc (to be conservative), which leads to a lower bound on the particle mass 
\beq
\lambda_{dB} = \frac{2 \pi}{m v} \lesssim 1 \, {\rm kpc} \Rightarrow m \gtrsim 5 \times 10^{-23} \, {\rm eV} \left(\frac{10^{-3}}{v}\right).
\eeq 
The lower bound is saturated in the fuzzy dark matter scenario~\cite{Hu:2000ke}, in which the quantum pressure of the ultralight bosons could support a core and solve the cusp-core problem in dwarf galaxies.

In this paper, I will focus on ultralight scalar dark matter with mass above $10^{-22} \, {\rm eV}$ and below eV. I will focus on the scenario with a \textit {repulsive} self-interaction between dark matter particles. The repulsive self-interaction could be the dominant source of pressure that supports a BEC. It is analogous to a BEC of dilute gas in traps. If the interaction between the atoms is attractive, the gas will collapse when the potential energy is above the kinetic energy. If the inter-atomic interaction is repulsive, a stable BEC could form (for a review of atomic BECs in traps, see~\cite{Dalfovo:1999zz}). In terms of the parametrization in Sec.~\ref{sec:par}, a repulsive self-interaction requires
\beq
\lambda_4 > \frac{5}{3} \lambda_3^2.
\label{eq:repulsive}
\eeq
There has been a lot of discussion of QCD axions forming BECs with galactic scale correlation length~\cite{Sikivie:2009qn, Erken:2011dz, Banik:2013rxa,Davidson:2013aba, Noumi:2013zga, Davidson:2014hfa}. But as argued in Ref~\cite{Guth:2014hsa}, QCD axions with a single $\cos$ potential have attractive self-interactions and the configuration with a long range correlation length is unstable against gravity. They are more likely to form local dense clumps such as Bose stars~\cite{Kolb:1993hw, Kolb:1993zz, Chavanis:2011zi, Chavanis:2011zm, Guth:2014hsa, Eby:2014fya}. 

Two colliding streams of pure dark matter condensate will not scatter from each other and dissipate energy if their relative velocity is below the sound speed of the condensate. Yet the relative velocity in the Bullet Cluster is estimated to be around 4700 km/s$\sim 0.016 c$~\cite{Randall:2007ph} or round 2860 km/s$\sim 0.01 c$~\cite{Springel:2007tu}. It is much larger than the sound speed one expects (one could estimate the sound speed by computing the circular velocity at the edge of a core with radius $\lesssim 1$ kpc, which is $\lesssim 10^{-3} c$). Thus in the Bullet Cluster, the colliding of streams will destroy the coherent state and one still needs to worry about the bound on dark matter self-interactions from the Bullet Cluster $\sigma/m \lesssim 1$ cm$^2$/g~\cite{Markevitch:2003at, Randall:2007ph}. In terms of parameters in Eq.~\ref{eq:scalar}, the self-interaction bound is translated into
\beq
\lambda_{4}^{\rm eff} \left(\frac{m}{f}\right)^2 \lesssim 10^{-11} \left(\frac{m}{\rm eV}\right)^{3/2},
\label{eq:selfinteractionbound}
\eeq
which shows that the quartic interaction has to be weak.

\section{Constructing a Reasonable Particle Physics Model}
\label{sec:model}

\subsection{Why is an Ultralight Repulsive Dark Matter Model Non-trivial?}

At first glance, one might think that it is trivial to write down a scalar dark matter model with a tiny mass term and a repulsive self-interaction, e.g., $V(\phi) = m^2/2 \phi^2 + (\lambda_4/4!) \phi^4$ in which $m$ and $\lambda_4$ are two uncorrelated free parameters with $\lambda_4 > 0$. Indeed this is the usual parametrization used in a lot of the studies on the astrophysical properties of ultralight scalar dark matter. Yet not every scalar potential is achievable and radiatively stable from the particle physics point of view. 
In particular, an (ultra-)light scalar is always dangerous in quantum field theory. Its potential has to be very flat and is subject to violent ultraviolet quantum corrections unless protected by an (almost exact) symmetry. What I will show is that it is highly non-trivial to have a ultralight repulsive dark matter model and it is more reasonable to adopt the parametrization in Eq.~\ref{eq:scalar}.

One way to (partially) protect a light scalar is to identify it as a Nambu-Goldstone boson with a decay constant $f$ resulting from the spontaneous breaking of a global symmetry. The Nambu-Goldstone boson itself enjoys a shift symmetry, i.e., invariance of its Lagrangian under $\phi(x) \to \phi(x) + c$ with $c$ a constant. The shift symmetry only allows derivative interactions of the scalar. A small explicit breaking of the shift symmetry will yield a mass and self-interactions of the scalar, making it a pseudo-Nambu-Goldstone Boson (pNGB). In this case, all the non-derivative scalar interactions will have their couplings proportional to $m^2$, the ``spurion" for the explicit breaking as demonstrated in Eq.~\ref{eq:scalar}. 
One well-known example is the QCD axion, a pNGB of a spontaneously broken $U(1)$ symmetry with a potential generated by the non-perturbative instanton effect
\beq
V(\phi)=\Lambda^4 \left(1 - \cos\left(\frac{\phi}{f}\right)\right),
\eeq
where $\Lambda$ is of order the QCD scale.\footnote{More precisely, the QCD axion potential is $V(\phi)=-m_\pi^2 f_\pi^2 \sqrt{1-\frac{4m_um_d}{(m_u+m_d)^2}\sin^2\left(\frac{\phi}{2f}\right)}$~\cite{DiVecchia:1980yfw}. The quartic coupling is $-0.35 \frac{m^2}{f^2}\phi^4$~\cite{diCortona:2015ldu}. } Expanding the potential in terms of $\phi/f$, one finds that $m^2 = \frac{\Lambda^2}{f}$ and the leading self-interaction, originating from the negative quartic coupling term, $-
\frac{m^2}{24f^2} \phi^4$, is attractive. In general, a pNGB from the breaking of a compact group has a periodic trigonometric potential and a single $\sin$ or $\cos$ potential always leads to an {\em attractive} self-interaction.

One might wonder whether a pNGB from breaking of a non-compact group could have a repulsive self-interaction, for example, a dilaton from the spontaneous breaking of scale invariance. Indeed, the dilaton potential is not periodic at all. Yet I will show that in calculable models, explicit breaking will generate a triple coupling of the dilaton, which leads to an effective {\it attractive} self-interaction in the NR limit! 
One might also wonder whether repulsive self-interactions could be achieved in ``monodromy" type potential: a cosine or sine potential modulated by a non-periodic function~\cite{Silverstein:2008sg}. Yet the monodromy potential locally is similar to a single trigonometric potential and still leads to an attractive self-interaction. 

Now one could see that it is highly non-trivial to construct an ultralight repulsive dark matter model. Yet it is not impossible. 
One way to obtain repulsive self-interaction is to have multiple trigonometric terms in a potential with specific coefficients. For instance, suppose the scalar potential has several cosine terms, some with negative coefficients while others have positive coefficients. If the cosine terms with negative coefficients contribute dominantly to the mass term while those with positive coefficients contribute dominantly to the quartic interaction, the mass term and quartic term in the scalar potential will have the same sign and the scalar has a repulsive self-interaction. This can only be achieved for specific sets of coefficients. I will show that this mechanism could be realized in an axion model from a 5D gauged $U(1)$ theory containing massive charged matter.

Another possible class of scalar dark matter models with repulsive self-interaction is the ``little dark matter" models. In this scenario, I propose that dark matter is a pNGB (inside a multiplet) from a collective symmetry breaking, which has been used in the little Higgs models to stabilize the electroweak scale at TeV~\cite{ArkaniHamed:2002qy, ArkaniHamed:2002qx}. The essence of collective symmetry breaking is to generate a positive quartic coupling for the pNGBs without a mass term at tree level. However, the quartic couplings in the little dark matter models are too big and much above the self-interaction bound in Eq.~\ref{eq:selfinteractionbound} for dark matter mass at or below the eV scale. It is still interesting as it could be a viable self-interacting MeV dark matter model.

Below in Sec.~\ref{sec:5Daxion}, I will present the model with an axion from a 5D gauged symmetry as a proof of concept that an ultralight scalar with repulsive self-interactions could exist. Then I will go through two classes of pNGB models to demonstrate that a large variety of models with scalar potentials completely different from the single $\cos$ or $\sin$ potential in minimal pNGB models still lead to attractive self-interactions (Sec.~\ref{sec:dilaton} and Sec.~\ref{sec:monodromy}). Finally I will discuss the little dark matter models in Sec.~\ref{sec:littledm} which gives a repulsive self-interaction but violates the self-interaction bound from the Bullet Cluster.

\subsection{A Working Example: Axion from a 5D Gauged $U(1)$}
\label{sec:5Daxion}
I will first show an axion model that could lead to repulsive self-interaction in the NR limit as a proof of concept. The axion considered here is not a QCD axion. 
In the model, the axion is a remnant of a five dimensional $U(1)$ gauge symmetry. In the 5D theory, the fifth dimension is compactified on a circle with radius $R$. The axion is identified as the gauge invariant Wilson loop of the fifth component of the gauge field, $A_5$, around the circle
\beq
\phi (x_\mu) &\equiv & f \oint  dx^5 A_5( x_\mu, x_5), \nonumber \\
f &\equiv & \frac{1}{2\pi Rg_4},
\eeq
where index $\mu$ labels the four dimensions we live in and $g_4^2 = g_5^2/(2\pi R)$ is the effective 4D gauge coupling
The 5D gauge symmetry $A_5 \to A_5 + \partial_5 c$ translates into a shift symmetry of $\phi$. At energies below $1/R$, one obtains an effective field theory for $\phi$ with the one-loop effective potential generated by 5D matter with charge $q$ and mass $m_5$~\cite{Hosotani:1983xw, Hatanaka:1998yp, Delgado:1998qr, Cheng:2002iz}
\beq
V(\phi) &=& -\frac{3(-1)^S}{4\pi^2}\frac{1}{(2\pi R)^4} \sum_{n=1}^{\infty} c_n e^{-2\pi n R m_5} \cos\left( \frac{n q\phi}{f}\right), \nonumber \\
c_n &=& \frac{(2\pi R m_5)^2}{3n^3}+ \frac{2\pi R m_5}{n^4} + \frac{1}{n^5},
\label{eq:5daxion}
\eeq
where $S=0$ for bosons and $S=1$ for fermions. This model has been discussed in the context of ``Extranatural Inflation''~\cite{ArkaniHamed:2003wu}. 
If $m_5\gg 1/(2\pi R)$, the higher order harmonic terms with $n>1$ will be suppressed by $e^{-2 \pi n R m_5}$. If $m_5 \ll 1/(2\pi R)$, the higher harmonics will be suppressed by $1/n^5$.

\begin{figure}[h]\begin{center}
\includegraphics[width=0.4\textwidth]{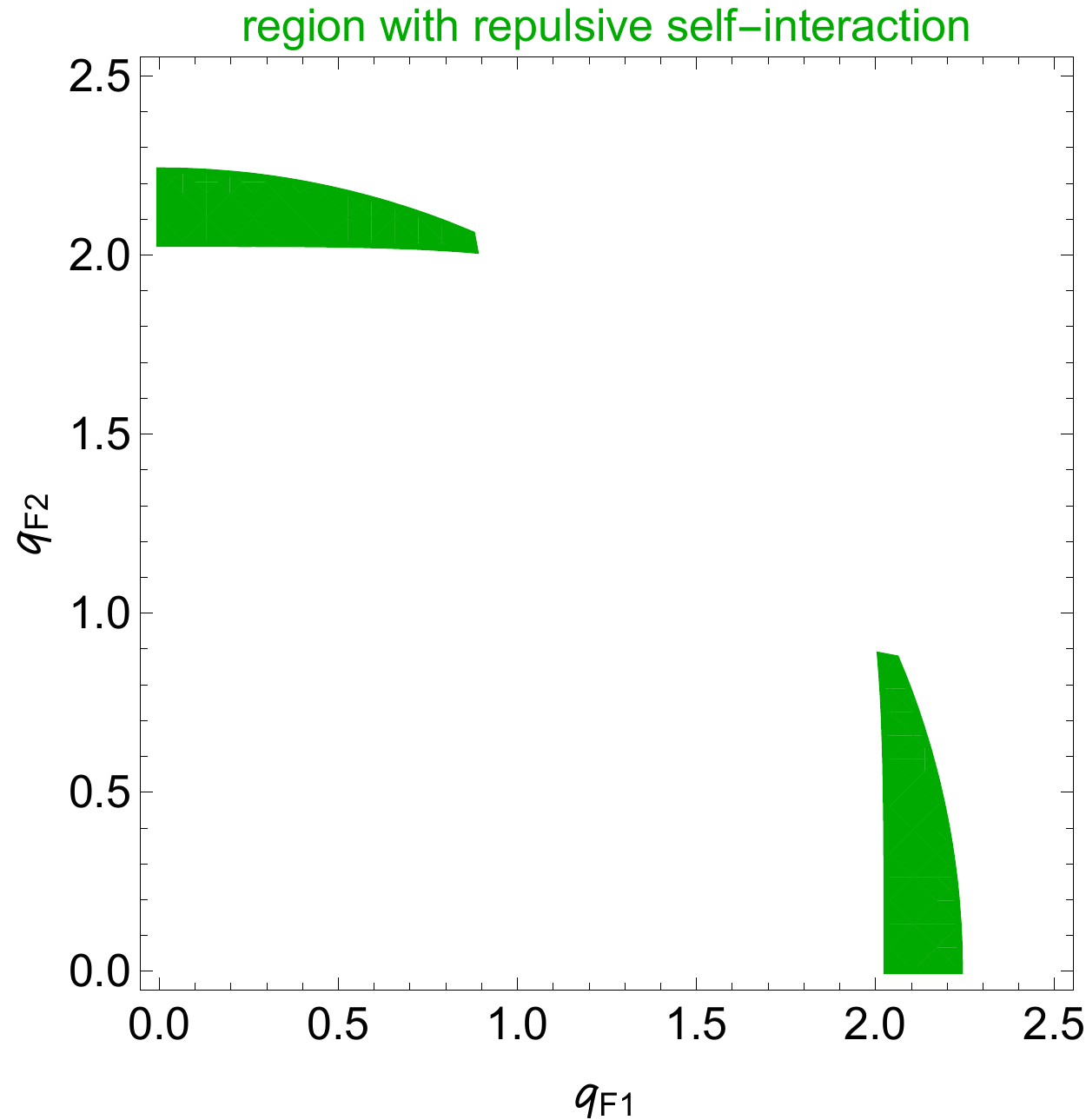}
\end{center}
\caption{Region (green) in the charge space $(q_{F1}, q_{F2})$ with repulsive self-interaction fixing $q_{B1} = 1, q_{B2} =2$. }
\label{fig:chargespace}
\end{figure}

One might wonder whether one could change the sign of the self-interaction by having both fermions and bosons contributing to the potential with opposite signs. Let's first consider a set of $n$ 5D bosons and $m$ 5D fermions with the same mass $m_5 \ll 1/(2\pi R)$. In this case, the higher harmonics with $n>1$ could be neglected. The bosons have a set of charges denoted by $\{q_{Bi}\}$ and the fermions have charges denoted by $\{q_{Fi}\}$. The effective axion potential contains
\beq
V(\phi) \supset - \Lambda^4\left(\sum_{i=1}^n  \cos\left(\frac{q_{Bi} \phi}{f}\right)-\sum_{i=1}^m   \cos\left(\frac{q_{Fi} \phi}{f}\right)\right),
\eeq
where $\Lambda$ is determined by $R$ and $m_5$. These two sums prefer different minima so at one minimum, the two sums will contribute to the potential oppositely. Expanding the potential around $\phi =0$, one gets 
\beq
V(\phi) \supset \frac{\Lambda^4}{2 f^2} \left(\sum_{i=1}^n q_{Bi}^2 - \sum_{i=1}^m q_{Fi}^2\right) \phi^2 + \frac{\Lambda^4}{24 f^4}\left (-\sum_{i=1}^n q_{Bi}^4 + \sum_{i=1}^m q_{Fi}^4 \right) \phi^4.
\eeq 
Requiring $\phi = 0$ to be a minimum\footnote{It is a local metastable minimum with a lifetime longer than the age of the Universe given $f\gg \Lambda$.} and to obtain a repulsive self-interaction, the following conditions have to be satisfied
\beq
\sum_{i=1}^n q_{Bi}^2 &>& \sum_{i=1}^m q_{Fi}^2 \\
\sum_{i=1}^n q_{Bi}^4 &<& \sum_{i=1}^m q_{Fi}^4.
\eeq
The two inequalities cannot be satisfied if either $n$ or $m$ is one. One needs at least two light fermions and two light scalars with charges chosen carefully to satisfy the two inequalities, for instance, $q_{B1} = 1, q_{B2} =2$ and $q_{F1} =2.1, q_{F2} = 0.6$. While satisfying the two inequalities is a tuning in the charge space as shown in Fig.~\ref{fig:chargespace}, it serves as a proof of concept that there exists light scalars with repulsive self-interaction. 

So far I have assumed all the 5D matter has the same mass $m_5 \ll 1/(2\pi R)$. One could also play with models with both different masses and charges. For instance, in the presence of a 5D boson with mass $2\pi m_B R =1$ and charge $q_B =1$ and a 5D fermion with mass $3.8<2 \pi m_F R<5.8$ and charge $q_F=2$, the effective axion quartic interaction is repulsive. Again only small islands of the parameter space in this setup allow for a repulsive self-interaction. 

\subsection{Dilaton}
\label{sec:dilaton}

Now I will discuss a model where a pNGB arises from spontaneous breaking of a non-compact global symmetry, the conformal symmetry. Although the potential is very different from that of QCD axion, it still leads to an attractive self-interaction in the NR limit. 

Consider a conformal sector in which the conformal symmetry is broken spontaneously and produces a dilaton. I will introduce a field $\chi(x)$ as the conformal compensator and the Goldston boson, i.e.~the dilaton, is defined as $\phi \equiv \chi - \langle \chi \rangle = \chi -f$. In addition, the conformal symmetry is broken explicitly by adding an operator ${\cal O}(x)$ with scaling dimension $\Delta_{\cal O} \neq 4$ to the Lagragian
\beq
{\cal L}_{\rm CFT} \to {\cal L}_{\rm CFT} + \lambda_{\cal O} {\cal O}(x).
\eeq
The pattern of symmetry breaking can be included in the effective field theory of dilaton by treating $\lambda_{\cal O}$ as a spurion. Using the spurion trick, one finds that the potential of $\chi$ is of form~\cite{Rattazzi:2000hs}
\beq
V(\chi)=\chi^4 \sum_{n=0}^{\infty} c_n(\lambda_{\cal O})\left(\frac{\chi}{f}\right)^{n(\Delta_{\cal O}-4)}, 
\eeq
where the dimensionless coefficients $c_n \sim \lambda_{\cal O}^n$, which is determined by the dynamics of the underlying field theory. By assumption, the dynamics should be that the potential is minimized at $\langle \chi \rangle = f$ with $m^2_\chi  > 0$.

It is in general difficult to calculate $c_n$'s in a strongly-coupled theory. Yet there are calculable cases with small explicit breaking and thus a small expansion parameter. One possibility is that the operator ${\cal O}$ is nearly marginal $\epsilon \equiv \left| \Delta_{\cal O}-4\right | \ll 1$ ($\lambda_{\cal O}$ is arbitrary). This is used in the 4D realization of the Goldberg-Wise mechanism~\cite{Goldberger:1999uk} to stabilize the extra-dimension in Randall-Sundrum model~\cite{Randall:1999ee}.  If this happens, the potential for the dilaton is, to the leading order of $\epsilon$, 
\beq
V(\phi) = \frac{1}{2} m^2 \phi^2 + \frac{5}{3!} \frac{m^2}{f} \phi^3+ \frac{11}{4!} \frac{m^2}{f^2} \phi^4+\cdots,
\eeq
where the mass absorbs $\epsilon$. The cubic interaction has been computed before in the context of a heavy dilaton faking the Higgs boson at a hadron collider in Ref~\cite{Goldberger:2008zz}. Unlike the axion case, the dilaton has a non-trivial cubic interaction which contributes negatively to the quartic interaction in the NR limit. The net quartic interaction in the NR limit, defined in Eq.~\ref{eq:effectivecoupling}, is
\beq
\lambda_4^{\rm eff} = -\frac{92}{3}.
\eeq

Another possibility of small explicit breaking is that $\lambda_{\cal O} \ll 1$ in unit of $f$. In this case, the dilaton potential is, to the leading order of $\lambda_{\cal O}$, 
\beq
V(\phi) = \frac{1}{2} m^2 \phi^2 + \frac{\Delta_{\cal O}+1}{3!} \frac{m^2}{f} \phi^3+ \frac{3-2\Delta_{\cal O}+\Delta_{\cal O}^2}{4!} \frac{m^2}{f^2} \phi^4+\cdots,
\eeq
where $m$ absorbs $\lambda_{\cal O}$. This leads to an effective quartic coupling in the NR limit
\beq
\lambda^{\rm eff}_4 =-\frac{2}{3}\Delta_{\cal O}^2-\frac{16}{3}\Delta_{\cal O}+ \frac{4}{3},
\eeq
which again is negative.\footnote{Mathematically, if $\Delta_{\cal O} < 0.24$, $\lambda^{\rm eff}_4 > 0$. Yet it doesn't make sense to talk about an operator with dimension less than 1 physically.}

To sum up, although the dilaton physics and potential are dramatically different from the QCD axion case, the dilaton's self-interaction is still {\bf attractive} in the NR limit!  

\subsection{Axion Monodromy}
\label{sec:monodromy}
Let's have some fun with the monodromy type potential. This type of potential is mostly discussed in the context of axion large field inflation and is probably the most sensible way to have the axion scan a field range above the Planck scale. The original monodromy model in Ref~\cite{Silverstein:2008sg} introduces a monodromy term generated by space-filling wrapped brane in addition to the usual instanton generated periodic potential
\beq
V(\phi)=\mu^3\sqrt{\ell^2+\phi^2} + \Lambda^4\left(1-\cos \left(\frac{\phi}{f}\right) \right),
\eeq
where $\mu, \ell$ are functions of compactification parameters in the underlying theory. Expanding around the minimum $\phi = 0$, one finds that the first term $\sqrt{\ell^2+\phi^2} \supset \phi^2/(2\ell)- \phi^4/(8 \ell^3)$ again contributes negatively to the quartic coupling resulting in an attractive self-interaction. Thus around the minimum, the self-interaction is always attractive no matter which term dominates.

\subsection{Collective Symmetry Breaking and Little Dark Matter}
\label{sec:littledm}

Another possible way to get repulsive self-interaction of the pNGB relies on more complicated symmetry breaking pattern to realize the collective symmetry breaking mechanism. For the collective symmetry breaking to work, one needs at least two sets of pNGBs with one set identified as the DM multiplet which will be denoted as $\phi$ and the other set denoted by $\xi$, which is heavier and could decay to two $\phi$'s. Both $\phi$ and $\xi$ could not be real singlets; otherwise, there will be dangerous tadpole terms generated at the radiative level~\cite{Schmaltz:2008vd}. They both transform non-trivially under the unbroken global symmetries. After the spontaneous breaking of a global symmetry (down to its subgroup), the potential of $\phi$ and $\xi$ is 
\beq
V(\phi, \xi) = \lambda_1 f^2 \left| \xi + \frac{[\phi \phi]}{f} + \cdots \right|^2 + \lambda_2 f^2 \left| \xi - \frac{[\phi \phi]}{f} + \cdots \right|^2,
\eeq
where $\lambda_1$ and $\lambda_2$ are order one dimensionless coefficients and the dots represent higher-order terms. $[\phi \phi]$ represent a possible contraction of two $\phi$ multiplets (e.g., $\phi^i \phi^j$ with $i, j = 1, 2$ if $\phi$ is a $SU(2)$ doublet). The two terms enjoy different shift symmetries
\beq
\phi &\to& \phi + c,  \\
\xi &\to& \xi - \frac{[\phi c]+[c \phi]}{f} \quad {\rm or} \quad \xi \to \xi + \frac{[\phi c]+[c \phi]}{f},
\eeq 
where $c$ is the constant shift (it is a constant vector or matrix since $\phi$ is a multiplet). Each individual term does not generate a potential for $\phi$ since one could always write each single term as merely a mass term for the heavy pNGB by a field redefinition. Yet the two terms together break the shift symmetries entirely and lead to an effective potential for $\phi$ after one integrates out the heavy pNGB $\xi$ with mass $(\lambda_1 + \lambda_2) f^2$ at tree level:
\beq
V_{\rm eff} (\phi) = \frac{4\lambda_1\lambda_2}{\lambda_1+ \lambda_2} [\phi \phi]^2 + \cdots \nonumber 
\eeq
The leading terms for $\phi$ is the quartic coupling and the coefficient is an order one positive number leading to a repulsive interaction. At one-loop level, integrating out $\xi$ generates a radiative mass term for $\phi$:
\beq
m_\phi^2 \sim \frac{\lambda_1^2+\lambda_2^2}{16\pi^2} f^2 
\eeq
up to an order one $\log$ factor which I didn't write out explicitly. In terms of the parameters in Eq.~\ref{eq:scalar}, in little dark matter models, 
\beq
\lambda_3 =0, \quad \lambda_4 \sim (4\pi)^2, \quad \frac{m}{f} \sim \frac{1}{4\pi}.
\eeq
This is the same dimensional analysis as in little Higgs models~\cite{ArkaniHamed:2002qy, ArkaniHamed:2002qx}. Unfortunately the self-interaction strength is too strong and violates the self-interaction bound in Eq.~\ref{eq:selfinteractionbound} for dark matter with mass below eV. 
Thus the collective symmetry breaking mechanism does not give us a viable ultralight repulsive dark matter model. Yet it could still satisfy the self-interaction bound if the dark matter mass is about or above 10 MeV and gives us a new class of self-interacting scalar dark matter model.

\section{Cosmological and Astrophysical Constraints}
\label{sec:cos}

In this section, I will discuss some cosmological and astrophysical constraints on ultralight repulsive dark matter. 
There have already been quite a few studies before on this topic (e.g. \cite{Riotto:2000kh, Boyle:2001du, Johnson:2008se, Arbey:2003sj, Boehmer:2007um, Harko:2011xw, Robles:2012uy, Dwornik:2013fra, Li:2014maa, Lora:2011yc, Harko:2011jy, RindlerDaller:2011kx, Chavanis:2012pd, Diez-Tejedor:2014naa, Chavanis:2011zm, Chavanis:2011zi, Chavanis:2011uv, Suarez:2015fga, Takeshi:2009cy, Fukuyama:2007sx, Fukuyama:2005jq}). What is new here is an interpretation in terms of the scalar potential parametrized by Eq.~\ref{eq:scalar} and Eq.~\ref{eq:effectivecoupling}, which manifest the implications for the underlying particle physics models.

\subsection{Early Universe Constraints}
\label{sec:earlyuniv}

In the early Universe, the ultralight dark matter starts to oscillate coherently around the minimum of the potential once the Hubble scale drops around its mass. The energy density of the coherent oscillation redshifts as that of ordinary cold and pressureless dark matter $\rho \propto a^{-3}$. 
Yet unlike the ordinary cold dark matter, the fluid of ultralight scalar dark matter are not pressureless due to small self-interactions. The dark matter fluid has a non-zero sound speed $c_s^2$ in contrast to the almost zero sound speed of ordinary cold dark matter. This could lead to modifications in the matter power spectrum. Below I'll calculate the sound speed and Jeans instability scale using the parametrization in Eq.~\ref{eq:scalar} and Eq.~\ref{eq:effectivecoupling}.

Consider the scalar field in an expanding Universe with the FRW metric $ds^2 = - dt^2 + a^2(t) d\vec{x}^2$. The perturbation of the metric could be parametrized as 
\beq
ds^2 = - (1+2\phi_N )dt^2 + (1-2\phi_N)a^2d\vec{x}^2,
\eeq
where $\phi_N$ is the metric perturbation and I ignore anisotropy.
To the leading order in the NR limit, the equations of motion for the scalar field $\psi$ (defined in Eq.~\ref{eq:nrpr}) and the metric perturbation $\phi_N$ are
\beq
&& i\frac{\partial_t(a^{3/2}  \psi)}{a^{3/2}}+  \frac{\nabla^2 \psi}{2ma^2} + \frac{\mu^2}{2m} \psi - \frac{\lambda^{\rm eff}_4}{8 f^2}|\psi|^2\psi -m \phi_N \psi= 0, \nonumber \\
&&\nabla^2 \phi_N = 4 \pi G a^2 (\delta \rho), 
\label{eq: eqofmottion}
\eeq
where $\delta \rho = \rho -\bar{\rho}\equiv m(|\psi|^2-\psi_c^2)/2$ is the perturbation of the energy density around a homogeneous background with energy density $\bar{\rho} \equiv m\psi_c^2/2$. 
I add in a chemical potential term $\mu^2 \psi/(2m)$ (equivalent to adding a $\mu^2 \phi^2/2$ term in the relativistic Lagrangian and $\mu^2 |\psi|^2/(2m)$ term in the NR Lagrangian) to account for a non-zero homogeneous background. The chemical potential $\mu$ is fixed by the energy density of the dark matter fluid, $\mu^2 = \lambda_4^{\rm eff} \bar{\rho}/(2f^2)$. The second equation is simply the Poisson equation in an expanding Universe and $\phi_N$ could be identified as the Newtonian gravitational potential.

The fluctuations around the homogeneous background can be parametrized as 
\beq 
\psi(a) =\frac{\psi_c^0+r }{a^{3/2}}e^{i \theta},
\eeq
where $\psi_c^0e^{i \theta}$ is related to the current dark matter density $\bar{\rho}_0 = m (\psi_c^{0})^2/2$. This parametrization reflects the fact that dark matter density $\propto |\psi|^2$ redshifts as $1/a^3$. 
Plugging the parametrization into Eq.~\ref{eq: eqofmottion}, the lineralized equation of motion for the radial mode $r$ in the momentum space can be reduced to 
\beq
\ddot{r}+2H\dot{r} + \left(\frac{k^2}{4m^2a^2}+\frac{\lambda^{\rm eff}_4 \bar{\rho}}{4 m^2f^2}\right) \frac{k^2 r}{a^2} = 4\pi G \bar{\rho}r,
\label{eq:pert}
\eeq 
where $k$ is the comoving wavenumber. This is also the linear density perturbation equation as the density perturbation $\delta \equiv \delta \rho/\bar{\rho} \propto r$. 
The second term on the left-hand side is the usual Hubble friction while the third term comes from various pressures.  The effective sound speed can be read off as 
\beq
c_s^2 =  \frac{k^2}{4m^2a^2}+ \frac{\lambda^{\rm eff}_4 \bar{\rho}}{4 m^2 f^2}.
\eeq
The first term is the sound speed from quantum pressure, which has already been computed in Ref~\cite{Hu:2000ke} while the second term originates from scalar self-interaction.

I will focus on the perturbation growth in the matter domination phase, starting from the matter-radiation equality at about $z_{\rm eq} \approx 3370$ to $z_\Lambda \approx 0.5$ when the vacuum energy takes over. Setting the sound speed to be zero, one obtains the well-known result that the dark matter perturbation grows linearly with the scale factor $a$ (or equivalently $t^{2/3}$) in the matter domination phase. 
For negligible self-interactions, one finds the usual Jeans instability scale 
\beq
k_J=\left( \frac{2m^2\bar{\rho} a^4}{M_p^2}\right)^{1/4} = 9 \sqrt{ \frac{m}{10^{-22}\, {\rm eV}}} \left(\frac{\bar{\rho}_0}{1.3 \times 10^{-6} \, {\rm GeV}/{\rm cm}^3}\right)^{1/4} \left(\frac{a}{a_{\rm eq}}\right)^{1/4} \, {\rm Mpc}^{-1},  
\eeq
where $\bar{\rho}_0$ is the current dark matter energy density and $M_p = 1/\sqrt{8\pi G} = 2.4 \times 10^{18}$ GeV is the reduced Planck scale. 
For length scale below the Jeans scale $L_J=2\pi/k_J$ or equivalently $k > k_J$, quantum pressure due to the uncertainty principle (an increase in momentum delocalizes the particles more) dominates. In this case, the density perturbation stops growing and oscillates with time. Since the comoving Jeans scale is almost scale invariant, perturbation growth below $k_J$ generates a sharp break in the matter power spectrum at about $k_J^{\rm eq}$, the comoving wavenumber at the matter-radiation equality~\cite{Kamionkowski:1999vp, Hu:2000ke}. The Jeans scale in this case is also roughly the de Broglie wavelength of the particle in the ground state. When dark matter mass is around $10^{-22}$ eV, more careful numerical studies show that the power spectrum drops at about 4.5 Mpc$^{-1}$ and could give reasonable fits to the data of dwarf spheroidal galaxies~\cite{Hu:2000ke, Marsh:2010wq, Marsh:2013ywa, Hlozek:2014lca, Marsh:2015wka}.

\begin{figure}[H]\begin{center}
\includegraphics[width=0.45\textwidth]{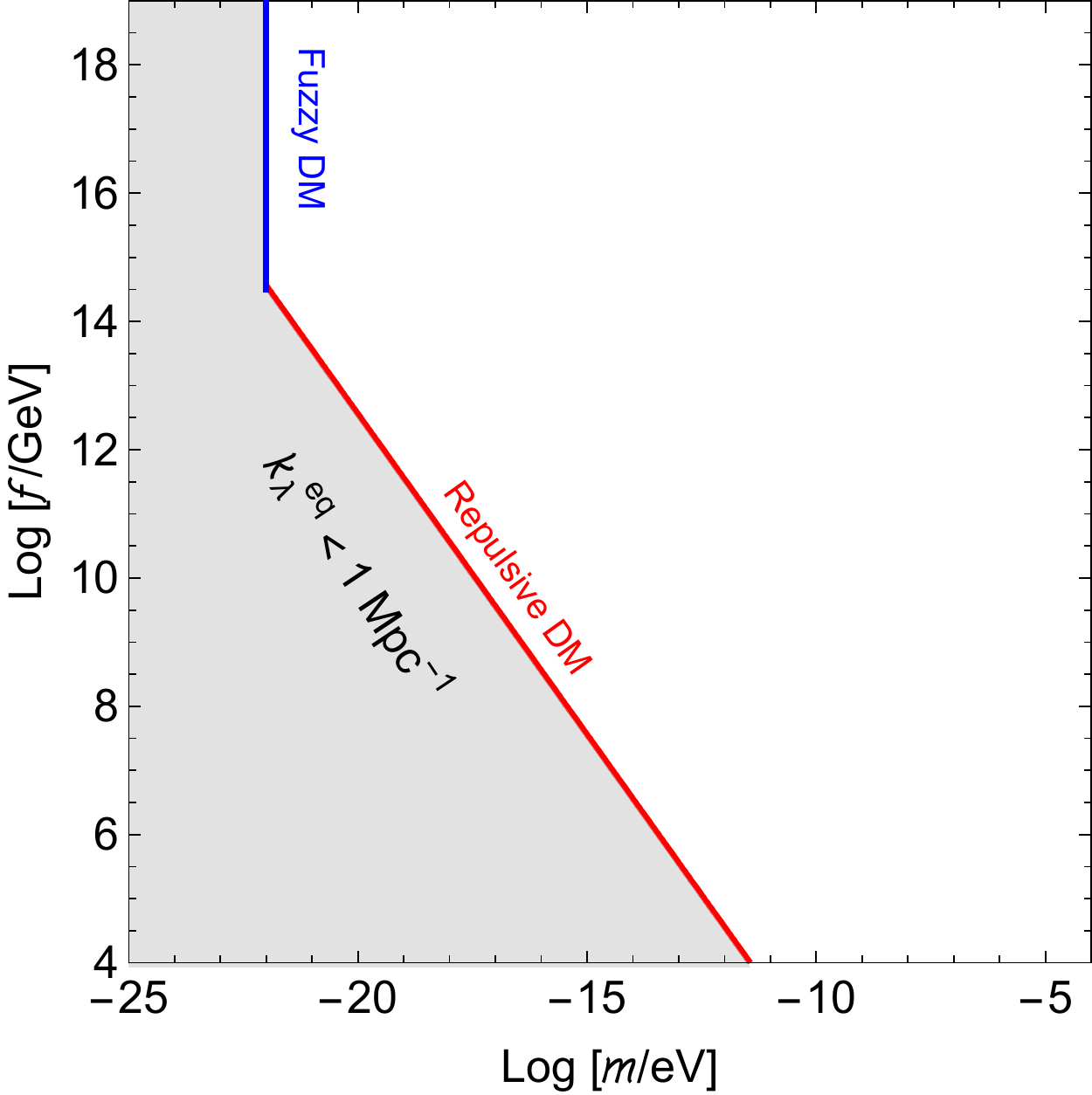}
\end{center}
\caption{Region in the $(m, f)$ plane with $k_\lambda^{\rm eq} > 1$ Mpc$^{-1}$ for $\lambda_4^{\rm eff} =1$ (above the shaded region). The vertical line at $m = 10^{-22}$ eV corresponds to the fuzzy dark matter case where the pressure from self-interaction is negligible compared to the quantum pressure. }
\label{fig:correlation}
\end{figure}

For $\lambda_4^{\rm eff}>0$, the repulsive self-interaction introduces another type of pressure (scaling as $k^2$) in addition to the quantum pressure (scaling as $k^4$). 
The repulsive self-interaction is the dominant source of pressure at smaller $k$. When it dominates, a new Jeans instability scale arises with $L_\lambda = 2 \pi/k_\lambda$ and 
\beq
k_{\lambda} = \sqrt{\frac{2}{\lambda_4^{\rm eff}}}\frac{m f a}{M_p} = 2.7\left( \frac{m}{10^{-20}\, {\rm eV}}\right)\left( \frac{f}{10^{13} \, {\rm GeV}}\right) \sqrt{\frac{1}{\lambda_4^{\rm eff}}}\left(\frac{a}{a_{\rm eq}}\right) \,{\rm Mpc}^{-1}.
\eeq
Below $L_\lambda$ (above $k_\lambda$ in the $k$ space), the repulsive interaction will suppress the growth of density perturbation and structure.\footnote{Similar analyses have been performed for general potentials of a real or complex scalar field with applications not only to dark matter but also to dark energy~\cite{Boyle:2001du, Johnson:2008se}.} Notice that the comoving wavenumber $k_\lambda$ scales as $a$ and equivalently the physical wavenumber is constant. It means that the instead of a sharp break in the matter power spectrum, the suppression of the perturbation will spread over several orders of magnitude in scale~\cite{Goodman:2000tg,Peebles:2000yy}. Without further detailed calculation, I will require $k_\lambda^{\rm eq} > 1$ Mpc$^{-1}$ as a very crude estimate to access the parameter space in the $(m, f)$ plane that could be consistent with the measurement of matter power spectrum. The result is shown in Fig.~\ref{fig:correlation}. In the figure, I also show a vertical blue line at $m = 10^{-22}$ eV corresponding to the fuzzy dark matter case where the pressure from self-interaction is negligible compared to the quantum pressure. The solid red line in the figure indicates the region where the repulsive self-interactions could have interesting consequences at smaller scale (next section) without modifying the large-scale structure.

\begin{figure}[h]\begin{center}
\includegraphics[width=0.33\textwidth]{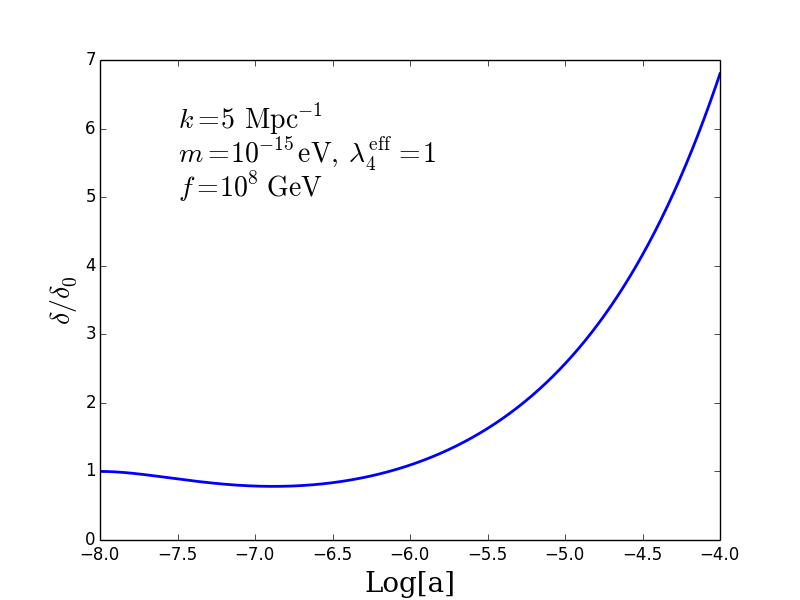}  \includegraphics[width=0.33\textwidth]{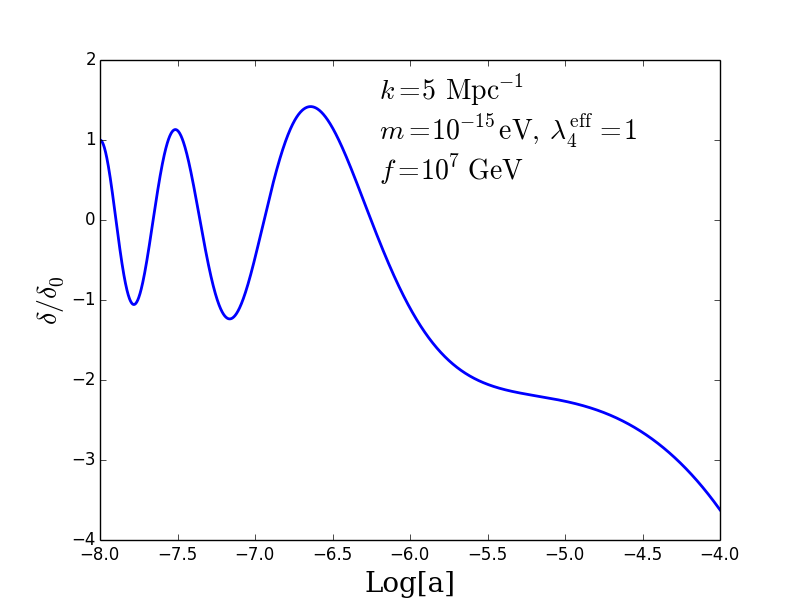}   \includegraphics[width=0.33\textwidth]{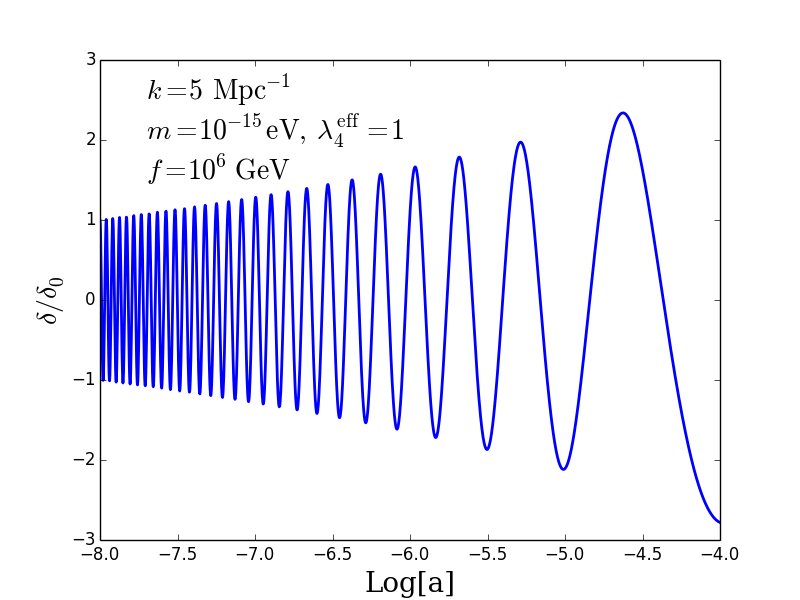}
\end{center}
\caption{Solutions to the dark matter density perturbation equation in Eq.~\ref{eq:pert} for difference choices of parameters. The comoving wavenumber is chosen to be 5 Mpc$^{-1}$. $m$ is fixed at $10^{-15}$ eV and $\lambda^{\rm eff}_4$ is chosen to 1. $f$ is $10^8$ GeV (left), $10^7$ GeV (middle) and $10^6$ GeV (right). 
}
\label{fig:solution}
\end{figure}

The perturbation equation, Eq.~\ref{eq:pert}, could be solved numerically. Results are shown in Fig.~\ref{fig:solution}. For a positive $\lambda^{\rm eff}_4$ and a small enough $f$ (the middle and right panels), the solution oscillates rapidly due to the new source of pressure. Yet the pressure redshifts with the scale factor $a$ and decreases with time so the amplitude of the oscillation grows. However, it grows much slower than the usual linear growth for pressureless cold dark matter.  The larger $f$ is, the smaller the pressure is and the oscillation is also reduced.

\subsection{Galactic Dynamics}

Now I want to consider possible effects of ultralight repulsive dark matter in galaxies. One motivation to consider ultralight dark matter with pressure (quantum pressure and/or pressure from repulsive self-interaction) is to address the cusp/core problem in dwarf galaxies. If the ultralight dark matter particles condense at the center of dwarf galaxies, the condensate is described more appropriately as a wave and each individual particle in the condensate is delocalized. Thus the density profile is more likely to have a core near the center. Unlike the linear analysis of dark matter perturbation in the early Universe, galaxies are highly non-linear systems and it is much more difficult to obtain analytic results. Strictly speaking, the way to check this possibility is through numerical simulations. Yet so far simulations have only been performed for the fuzzy dark matter scenario (corresponding to the parametrization in Eq.~\ref{eq:scalar} with a large $f$ and thus a negligible self-interaction and effectively one free parameter $m$ left)~\cite{Schive:2014dra, Schive:2014hza}. It is found in the numerical simulations that indeed for $m \approx 10^{-22}$ eV, a core with radius $\sim$ kpc is formed at the center, which could be well fit by a stable soliton solution $\rho_s(r) \propto 1/(1+c_0 r^2)^8$ with $c_0$ a constant. Yet the outer density profile is similar to the Navarro-Frenk-White (NFW) profile of the standard cold dark matter~\cite{Navarro:1995iw}, which scales as $1/r^3$. For the case with a non-negligible repulsive self-interaction, I will review a rough estimate of possible dark matter core profile~\cite{Tkachev:1986tr, Goodman:2000tg} and perform a fit with a subset of dwarf spheroidal galaxy data. 

For a spherically symmetric fluid in hydrostatic equilibrium, 
\beq
\frac{dP(r)}{dr} = - \frac{G_N M(r)}{r^2} \rho(r),
\eeq
where $P(r)$ is the pressure as a function of radius $r$ and $M(r)$ is the mass enclosed in the sphere with radius $r$. The equation of state can be inferred from $dP/d\rho = c_s^2$
\beq
P(r)&=& K \rho(r)^2, \\
K&\equiv & \frac{\lambda_4^{\rm eff} }{8 m^2f^2}.
\label{eq:eos}
\eeq
This is the equation of state for a polytropic gas, $P(r)=K \rho(r)^\gamma$ with index $\gamma=2$. Combining the equations above and the continuity equation $dM(r)/dr = 4 \pi r^2 \rho(r)$, one finds an analytic solution for the dark matter core~\cite{Tkachev:1986tr, Goodman:2000tg}  
\beq
\rho (r) &=& \rho_0 \frac{\sin{(r/\alpha)}}{(r/\alpha)}, \\
\alpha &=& \sqrt{\frac{K}{2\pi G}},
\label{eq:coresolution}
\eeq
where $\rho_0$ is the density at the center $r=0$. 
Notice that this solution is only valid up to $r_c = \pi \alpha$ where the density is zero. 

It should be emphasized that the derivation above should only be taken as a reasonable crude guess rather than a robust predication. 
First it could be possible that the core solution will transit into an NFW-like outer region at  a smaller radius $r < r_c$ (analogous to what is observed in the simulations of fuzzy dark matter~\cite{Schive:2014dra, Schive:2014hza}).  In addition, the profile could be modified at non-zero temperatures (strictly speaking, the equation of state in Eq.~\ref{eq:eos} only holds at zero temperature) or by baryons and vary from galaxy to galaxy. Anyway, I will take Eq.~\ref{eq:coresolution} as the dark matter profile and apply it to analyze a subset of dwarf spheroidal data below. 

\begin{figure}[H]\begin{center}
\includegraphics[width=0.45\textwidth]{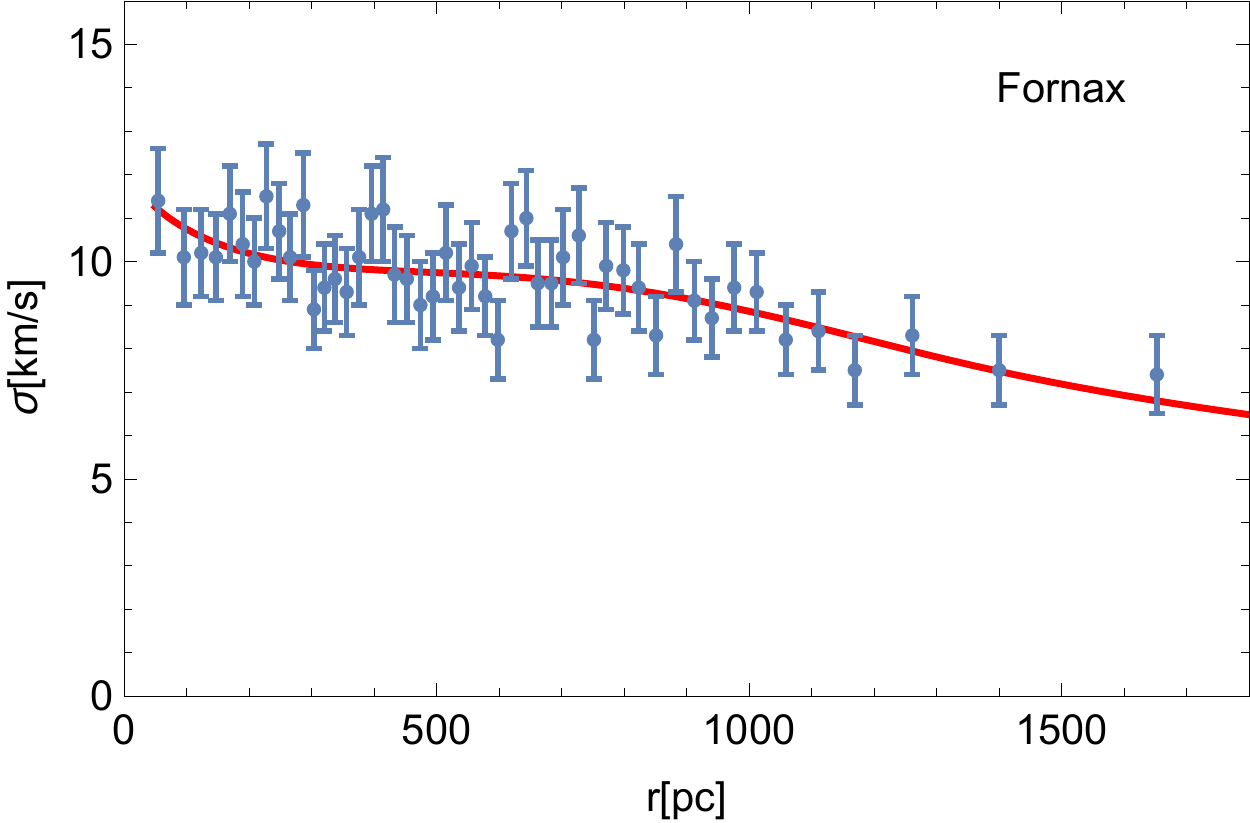} \quad  \includegraphics[width=0.35\textwidth]{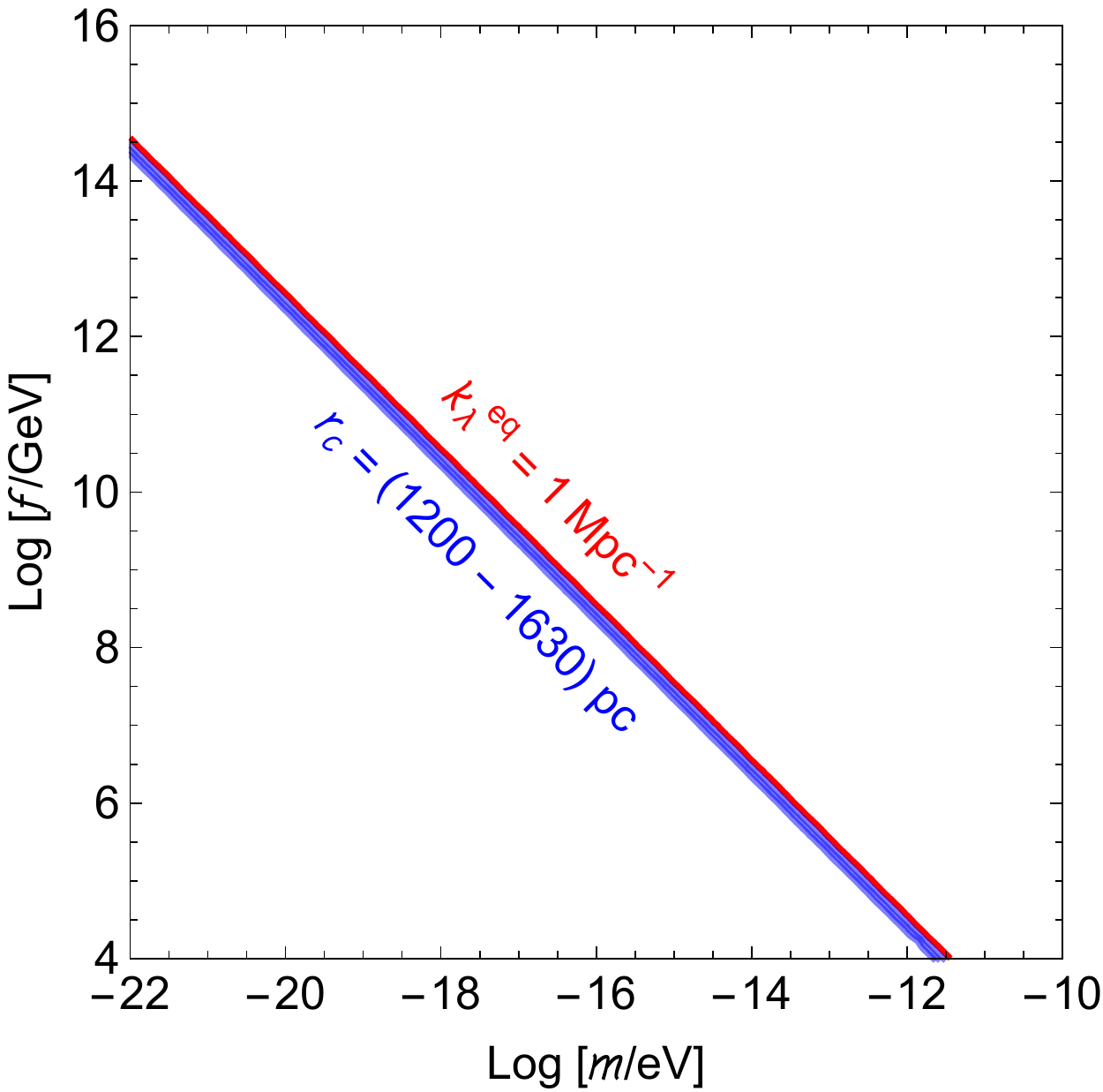}
\end{center}
\caption{Left: Best fit of the line-of-sight velocity dispersion as a function of distance from the center in Fornax assuming a dark matter density profile in Eq.~\ref{eq:coresolution}. Data points are from Ref~\cite{Walker:2009zp}. Right: the region in the $(m, f)$ plane with $r_c$ is in the range $(1200 - 1630)$pc (blue), which is allowed by the Fornax line-of-sight velocity dispersion data at 68\% C.L.  The red line gives $k_\lambda^{\rm eq} = 1$ Mpc$^{-1}$, the rough boundary below which the repulsion could modify the linear part of the matter power spectrum. }
\label{fig:profiles}
\end{figure}

There are eight classical Milky Way's dwarf spheroidal galaxies with high quality data sets of stellar kinematics. One important observable in the analysis of these dwarf galaxies is the line-of-sight velocity dispersion as a function of the projected radius. The definition and the measured data could be found in Ref~\cite{Walker:2009zp}. Below I will examine Fornax as a benchmark. I calculate the line-of-sight velocity dispersion assuming the core profile in Eq.~\ref{eq:coresolution} and perform a $\chi^2$ fit using the results in~\cite{Walker:2009zp}.\footnote{I thank Matt Walker for providing the data for Fig. 1 of Ref~\cite{Walker:2009zp}.} In the fit, I varied three parameters $\rho_0$, $r_c$ and $\beta(r) \equiv 1- \bar{v_{\theta}^2}/\bar{v_{r}^2}$ which describes the stellar anisotropy. 
The best fit projected velocity dispersion is presented in the left panel of Fig.~\ref{fig:profiles}. The $\chi_{\rm min}^2$/d.o.f of the best fit $\approx 0.6$. At 68\% C.L., the allowed $r_c$ is in the range $(1200 - 1630)$pc with the best fit at $r_c = 1400$ pc. This roughly agrees with the result in Ref.~\cite{Diez-Tejedor:2014naa}. The region in the $(m, f)$ plane giving rise to the allowed $r_c$ at 68\% C.L is presented in the right panel of Fig. \ref{fig:profiles}. It is almost on top of the contour with $k_\lambda^{\rm eq} = 1$ Mpc$^{-1}$, the rough boundary below which the repulsion could modify the linear regime in the matter power spectrum.

It has been argued that if the bosons reach thermal equilibrium, isothermal spherical halo model indicates too sharp a drop in density at the edge of BEC and could be ruled out by the velocity distribution~\cite{Slepian:2011ev}. In addition, a scaling relation is observed that the central density times the core radius is roughly a constant, which holds for all dwarf galaxies~\cite{Kormendy:2014ova, Burkert:2015vla}. To address the first potential constraint, one needs to check whether the transition from the core to the outer region is described by an isothermal model and to address the second observation, one needs to determine the location of the boundary between the core and outer region. These two issues are tied up and difficult to answer without numerical works. It is also pointed out in Ref.~\cite{Diez-Tejedor:2014naa} that the parameter space preferred by dwarf spheroidal galaxies is not consistent with that by low-surface-brightness galaxies and bigger galaxies~\cite{Arbey:2003sj, Boehmer:2007um, Harko:2011xw, Robles:2012uy, Dwornik:2013fra, Li:2014maa}.  The latter classes of galaxies prefer larger core radius of order 10 kpc, which is also in some tension with the matter power spectrum constraints. This raises the question how the bosons behave in different galaxies. I will leave these open questions for future investigations.

\section{Outlook and Future Directions}
\label{sec:con}

This paper aims at constructing reasonable particle physics models of ultralight dark matter with repulsive self-interactions. This class of dark matter scenarios may have interesting astrophysical consequences at the galactic scale such as solving the cusp/core problem (or explaining MOND phenomenology for special types of self-interactions). I discuss the difficulties in obtaining a viable model and list some illustrative examples demonstrating different challenges. Though difficult, there is still a working model of axions from a 5D gauged $U(1)$ with charged matter satisfying certain constraints in the charge/mass space which serves as a proof of concept. I discuss the cosmological constraints on the parametrization motived by particle physics considerations and possible galactic dynamics. The work is only a first step towards combining particle physics and cosmology/astrophysics for a more thorough investigation of the self-interactions of ultralight scalar dark matter and their possible consequences. More studies remain to be done. Here I will briefly outline some important directions for future work.

\textbf {Relic abundance.} The coherent oscillation of ultralight scalar dark matter could store energy and redshifts as ordinary matter. Yet in the parameter space that leads to interesting galactic dynamics, coherent oscillation with amplitude $f$ does not necessarily give the observed dark matter relic abundance. Yet it is possible that a non-standard cosmology (i.e., adding contribution from cosmic string decays~\cite{Davis:1986xc} or non-thermal cosmologies along the line proposed in Ref~\cite{Randall:2015xza}) may lead to the observed relic abundance.
Indeed ultralight scalar dark matter I considered in this paper is more motived by its potential interesting phenomenology rather than an elegant explanation of relic abundance. Nevertheless, it is useful to work out different cosmological histories compatible with data and check whether there could be other interesting observables. 

\textbf {Multi-component dark matter.} As discussed in the context of self-interacting dark matter, even a subdominant component of dark matter could lead to dramatically different phenomenology~\cite{Fan:2013tia, Fan:2013yva}. 
If the ultralight dark matter with repulsive self-interaction is only one component of dark matter, the cosmological constraints may be relaxed and the galactic dynamics could be very different too. It will be useful to revisit the cosmological and astrophysical properties including the fraction of ultralight dark matter in the total amount of dark matter as a free parameter.

\textbf {Dark atoms.} Another possibility I haven't discussed is that ultralight dark matter is a weakly-bound state such as a dark atom. Then dark BEC is exactly analogous to that in the cold atom system. 
Consider dark atoms made of a light fermion (i.e., a dark electron $e_D$ with mass $m_{e_D}$) and a heavier fermion (i.e., a dark proton $p_D$ with mass $m_{p_D}$). The necessary requirement for the formation of a BEC is then the de Broglie wavelength of the electron is larger than the inter-atomic spacing
\beq
\frac{m_{e_D}}{\rm eV} \left({\frac{m_{p_D}}{10\,{\rm eV}}}\right)^{1/3} < 1.6 \left(\frac{10^{-3}}{v}\right) \left(\frac{\rho}{1.3 \times 10^{-6}\, {\rm GeV}/{\rm cm}^3}\right)^{1/3}.
\eeq 
The big problem here is the existences of the light degrees of freedom such as dark photon. It remains to be seen whether there is an ultralight dark atom model with a viable cosmology. 

\textbf{Strongly-coupled NR system.} In this work, I assumed both the relativistic and NR Lagrangians are perturbative. In condensed matter physics, NR Lagrangians of strongly-coupled system have been constructed using the tool of effective field theory~\cite{Son:2005rv}. It has been used by Refs \cite{Berezhiani:2015pia, Berezhiani:2015bqa} to explain MOND phenomenology. A strongly-coupled NR Lagrangian may still be UV completed into a perturbative relativistic Lagrangian though it is difficult to implement the matching calculation. Yet it is still of great interest to derive consequences of different NR Lagrangians. 

\textbf{Density profiles and compatability with dwarf data.} I assume the simplest core profile to fit the data of dwarf spheroidal galaxies. For a better fit, one could take the same strategy as in Ref \cite{Marsh:2015wka} for fuzzy dark matter assuming a central core and an NFW profile at large radii and leaving the core radius as a free parameter. Fully addressing whether the ultralight dark matter could solve the cusp/core problem and explain the scaling relation in dwarf galaxies calls for a numerical simulation similar to Ref~\cite{Schive:2014dra, Schive:2014hza} but with both the mass and the interaction strength as free parameters.

\textbf{Other astrophysical probes.} It has been shown that gravity wave detection such as advanced LIGO may detect ultralight dark matter such as QCD axion with GUT scale decay constant through superradiance of stellar black holes~\cite{Arvanitaki:2014wva}. It will be interesting to use the LIGO data to test the broad parameter space of ultralight dark matter scenarios with repulsive self-interactions. Other possibilities include 21cm power spectrum~\cite{Marsh:2015daa} and isocurvature~\cite{Marsh:2015xka}. 

\textbf {New directions of dark matter detection.} While ultralight dark matter is way outside the sensitivity range of direct detection, it is still possible to be detected directly depending on its possible interactions with the standard model particles. There have been promising proposals for ultralight axions using solid-state NMR-based experiments~\cite{Budker:2013hfa} and resonant and broadband readout circuits~\cite{Kahn:2016aff} and for ultralight dilaton and moduli dark matter such as dedicated resonant-mass detectors and atomic clocks~\cite{Arvanitaki:2014faa, Arvanitaki:2015iga}. Interesting limits have already been derived on dark matter couplings to the standard model through non-observation of variations in the fundamental constants~\cite{Stadnik:2013raa, Stadnik:2014tta, Stadnik:2015kia}. It could be useful to develop more search strategies to cover the full parameter space of ultralight dark matter, in particular, the region with non-trivial astrophysical phenomenology. 

Finally I want to comment that the paper focuses on scalar dark matter. Another interesting possibility to investigate is the fermionic dark matter with mass around 10 - 100 eV and pressure arising from Pauli exclusion principle~\cite{Destri:2013pt, Domcke:2014kla, future}.

\section*{Acknowledgement}

I thank Prateek Agrawal, Francis-Yan Cyr-Racine, Savvas Koushiappas, David J.E. Marsh, Zackary Slepian, Matt Reece, Ben Safdi, Jakub Scholtz, Jessie Shelton, Jesse Thaler, Matt Walker for useful discussions and correspondences. I thank Harvard Center for the Fundamental Laws of Nature for hospitality during completion of this paper.

\bibliography{ref}
\bibliographystyle{utphys}
\end{document}